\documentclass[12pt, oneside]{article}   	
\usepackage{graphicx}				
\usepackage{amssymb}
\usepackage{amsmath}
\usepackage{hyperref}
\usepackage{xurl}

\title{Forecasting Extreme High Summer Temperatures in Paris and Cairo Using Gradient Boosting and Conformal Prediction Regions}
\author{Richard Berk \\ University of Pennsylvania}

\begin{document}
\maketitle

\begin{abstract}
In this paper, gradient boosting is used to forecast the Q(.95) values of air temperature and the Steadman Heat Index. Paris, France during late the spring and summer months is the major focus. Predictors and responses are drawn from the Paris-Montsouris weather station for the years 2018 through 2024. Q(.95) values are used because of interest in summer heat that is statistically rare and extreme. The data are curated as a multiple time series for each year. Predictors include seven routinely collected indicators of weather conditions. They each are lagged by 14 days such that temperature and heat index forecasts are provided two weeks in advance. Forecasting uncertainty is addressed with conformal prediction regions. Forecasting accuracy is promising. Cairo, Egypt is a second location using data  from the weather station at the Cairo Internal Airport over the same years and months. Cairo is a more challenging setting for temperature forecasting because its desert climate can create abrupt and erratic temperature changes. Yet, there is some progress forecasting record-setting hot days. 
\end{abstract}


\pagebreak
\section{Introduction}

Heat waves and their consequences have been a major research focus of climate science for over a decade (Perkins, 2015; Pitcar et al., 2019; Marx et al., 2020; Klingh\"ofer et al., 2023). With few exceptions (Russo et al., 2014; Jaque-Dumas, 2022; Khantana et al., 2024), heat wave forecasting has not been a guiding narrative. Description and explanation have  dominated (Mann, 2018; Mckinnon and Simpson, 2022; Petoukhov et al., 2022; Li et al, 2024; Leeper et al., 2025). Practical challenges to heat wave forecasting have included a substantial reliance on simulation output that can be ill-suited for studying rare and extreme events, and uncertainty quantification that Gettleman and Hood (2017) attribute to (1) model uncertainty, (2) scenario uncertainty, and (3) initial condition uncertainty. But, perhaps the most obvious obstacle is heat wave definitions that often lack precision and firm scientific justification. Perkins and Alexander (2013) provide a succinct assessment: ``Despite their adverse impacts, definitions and measurements of heat waves are ambiguous and inconsistent, generally being endemic to only the group affected, or the respective study reporting the analysis.'' \footnote
{
There are also legitimate differences in heat wave definitions that depend on location. For example, heat waves in China can be characterized by avoidable mortality that varies by region and can lead to lengthy heat waves when defined in that manner (Liu et al., 2021). East Asian countries can experience prolonged high-pressure systems, especially during the summer monsoon season, that can lead to extended periods of high temperatures. Desert locations often have very short but intense heat waves that will be addressed later.
} 
 
In this paper, we focus instead on forecasts of unusually high summer temperatures that sometimes can shape media characterization of heat waves (Hulme et al., 2008; Hopke, 2025). High temperatures may be newsworthy, but media heat wave accounts should not be confused with confirmed scientific facts. Forecasting can suffer. 

Our data-driven forecasts use data from a Paris weather station and a Cairo weather station for the years 2018 through 2024. Multiple time series observations are analyzed with gradient boosting and conformal prediction regions. We draw on the work of Berk and his colleagues (Berk et al., 2024; Berk and Braverman, 2025), who have used remotely sensed AIRS data to lay a foundation for our temperature forecasts. 

In section 2, the data and statistical procedures are described. Section 3 summarizes the Paris results. Section 4 illustrates possible difficulties that can follow when trying to forecast extreme high temperatures in some desert climates. Cairo, Egypt provides an illustration. Section 5, turns to policy implications. A general discussion of further steps is addressed in section 6. Section 7 offers some conclusions.

\section{The Data and Statistical Methods}

Hourly weather data were obtained from the Paris-Montsouris Weather Station using software in R that NOAA provides.\footnote
{
\texttt{paris\_data <- importNOAA(code = ``071560-99999'', year = 2018:2024)} reads and downloads Integrated Surface Data (ISD) from NOAA. Once the data were downloaded, R packages tidyverse, zoo, and lubridate are used to construct appropriate data frames. For the figures, ggplot2  is used but most of the statistical computing needed for the figures is done outside of ggplot2.  As of May 23rd, 2025, the downloading of the data was intermittently operational.
}
For 2018 through 2024, days are the temporal units. All measures used were obtained at 2pm solar time to capture the warmest parts of each day; there is one set of 2pm variables for each day. 

On any given day, the highest temperatures may not be measured at 2pm solar time. But, choosing a single time supports longitudinal compatibility, and temperatures at 2 PM will generally be a good indicator of high daily temperatures. The maximum daily temperature can be another option, but confounds time of day with temperature. High temperatures at night can have different consequences from high temperatures during the day. For public and ecosystem health, rising night-time temperatures may be of greater concern (Peng et al., 2013; He et al., 2022) but day-time temperatures seem to attract the most publicity (Graff, 2025). The data include night-time temperatures that are part of future analysis.  

The data are further filtered to include only March through August, months when the warmer temperatures are usually realized. March is not included because high temperature values are expected. March is included so that 14 day lagged values for the predictors are available for forecasting temperatures starting in early April. 

The main response variable is ``air temperature.'' For supplementary analyses, we compute as a response the Steadman relative humidity heat index. Predictors lagged by 14 days include (1) wind direction in degrees from true north, (2) wind speed in meters per second,  (3) air temperature in degrees celsius, (4)  atmospheric pressure hectopascals (hPa) (5) visibility in meters, (6) dew point in degrees celsius, and (7) relative humidity in percent units.  Several multiple time series are constructed separately for each year's relevant months and used in a manner explained shortly. 

A credible analysis requires training data and test data. Because of likely temporal dependence in how the data were realized, popular methods to construct legitimate test data using probability sampling are not a realistic option. Random sampling would debase any temporal structure. Conventional cross-validation, which also depends on random sampling, is also inappropriate. 

Hyndman and Athanasopoulos (2021: chapter 5) suggest using as test data a time series that begins later than the training data, coupled with a special kind of cross-validation that then might be valid in some settings. However, that is not practical here. Over the course of the summer and into the fall, a time series of air temperatures is not likely to be strongly stationary. For example, it is usually warmest mid to late summer and then cools. 

As an alternative, our test data will come from an adjacent year, but for the same warmer weather months. The issues can be subtle. For example, test data for the summer of 2020 could be data either from the summer 2019 or the summer 2021. Such test data would be treated as a product of random variables realized from the same underlying physics in the same months from one year to the next (Stull, 2017: chapter 3). A credible case for weak stationarity then might be made from covariance stationarity alone. By conditioning on month, the levels of the random variables could change over proximate years, but not their relationships with one another. This allows for daily temperature change in level such as might be caused by El Ni\~no events and others. We will see how this play out shortly.

The multiple time series data will be analyzed using the \textit{gbm} procedure in R written by Greg Ridgeway. It is a form of gradient boosting (Friedman, 2001; 2002) that includes a quantile loss function written by Brian Kriegler. In principle, any conditional quantile loss can be implemented in a nonparametric fashion within the gradient boosting structure. The Q(.95) value will be central because the 95th quantile helps to capture rare, extreme values. 

Forecasts are provided within conformal prediction regions. A complication is that the temporal dependence in the multiple time series data violates the requisite exchangeability (Chernozhukov et al., 2018).  However, if the residuals from the boosting analysis are not temporally dependent, the exchangeability assumptions can still apply (Berk et al., 2025). That alternative will be examined.\footnote
{
The option of using a recurrent neural network variants was not considered because of their well-known difficulties in training (e.g., ``vanishing" gradients). The more recent option of transformer time series analysis was also not considered. There  is not yet much practical experience with transformer applications to time series data. Moreover, as Ahmed and colleagues note (2023); ``... the self-attention mechanism has high computational complexity and memory requirements hampering long sequence modeling.'' It is also unclear if either neural network approach can do better than well-established and widely used supervised learning. 
}

In these analyses, the test data plays a vital but circumscribed role. Over iterations, gradient boosting will in general fit the training data better. At some point, training data overfitting is introduced, and there is no apparent stopping rule. This is why test data or some related approach are required. Over iterations, gradient boosting will in general fit the test data better as well, but at some point, further iterations degrade the fit. A common stopping rule is to halt the fitting process when the test data are fit as well as the data and algorithm can achieve. Conventional overfitting is precluded.\footnote
{
Under certain circumstance, a phenomenon called ``double descent''  can materialize. Over a very large number of iterations,  there will be with test data an early stopping solution after which the fit degrades, only to start improving again later (Belkin et al., 2019).  The issues are beyond the scope of this paper and seem to apply primarily to machine learning classifiers. Suffice it to say that as many as 20,000 iterations were run, and there was never more than one recommended sweet spot.
}

After the Paris analysis was completed, data were collected for Cairo from the weather station located at the Cairo International Airport. The choice was shaped by a desire to forecast rare and extreme high temperatures in a climate setting that substantially differs from the Paris climate. Paris has a temperate oceanic climate whereas Cairo has a hot desert climate. Winds from the Atlantic ocean bring moisture and moderating temperatures to Paris. Cairo's Khamsin winds, a bit like the Santa Ana winds in Los Angeles, bring hot, dry, dusty air for the nearby desert.  Very rapid temperature increases can result. The data collection, data curating, and data analyses are effectively identical in the two locations. 

\section{Paris Results}

Three pairs of years are analyzed: (1) 2019 as training data and 2018 as test data; (2) 2020 as training data and 2021 as test data; (2) 2022 as training data and 2023 as test data. Note that sometimes the test data precede the training data, and sometimes the test data follow the training data. We know of no \textit{a priori} method to scientifically justify the training and test data order. To analyze the data first and then choose which year will be used for the training data and which year will be used for the test data is to encourage data snooping. The guiding principle simply is to use adjacent years to arrive at training data and test data that will likely be structurally most comparable. 

\subsection{Years 2018 as Test Data  and 2019 as Training Data}

\begin{figure}[htbp]
\begin{center}
\includegraphics[width=3.5in]{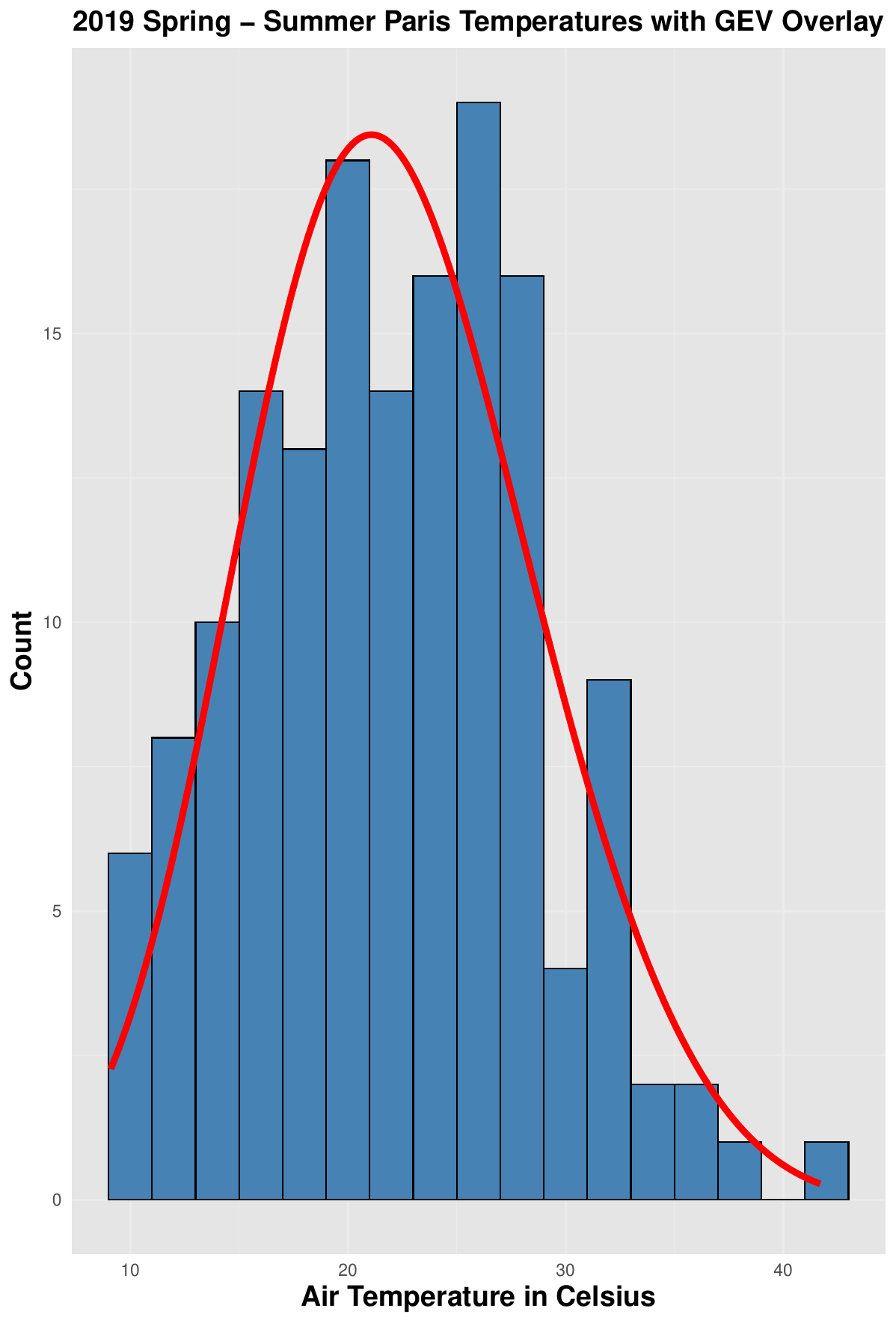}
\caption{Histogram of the Paris Daily 2 PM Air temperature. s in Celsius for March through August in 2019 with a generalized extreme value distribution overlay. To add context, 40\textdegree{}C is 104\textdegree{}F.
. (N = 153)}
\label{fig:hist2019}
\end{center}
\end{figure}

Figure~\ref{fig:hist2019} shows a histogram for the daily  2pm air temperature in the greater Paris area for March through August in 2019. A fitted generalized extreme value distribution is overlaid that seems to provide a reasonable summary. The distribution has a substantial right tail representing higher temperatures. The right tail includes some very hot days. 

\subsubsection{Fit Quality}

Figure~\ref{fig:hist2019} may suggest that one might apply a form of linear regression for extreme values distributions (Coles, 2001). There several problems with this approach. Figure~\ref{fig:hist2019} displays the marginal distribution for air temperature while our interests center on the conditional distributions that depend on the seven predictors listed earlier. There is no reason to expect the that conditional distributions are well approximated by a generalized extreme value density. The same concerns apply to the residuals from a linear regression model. More important, we aim to forecast extreme and rare high temperatures often associated with heat waves. This calls not for good estimates of conditional mean temperatures but for good estimates of very high conditional quantiles. The goal is not to estimate what is typical but what is atypically high.  

\begin{figure}[htbp]
\begin{center}
\includegraphics[width=4in]{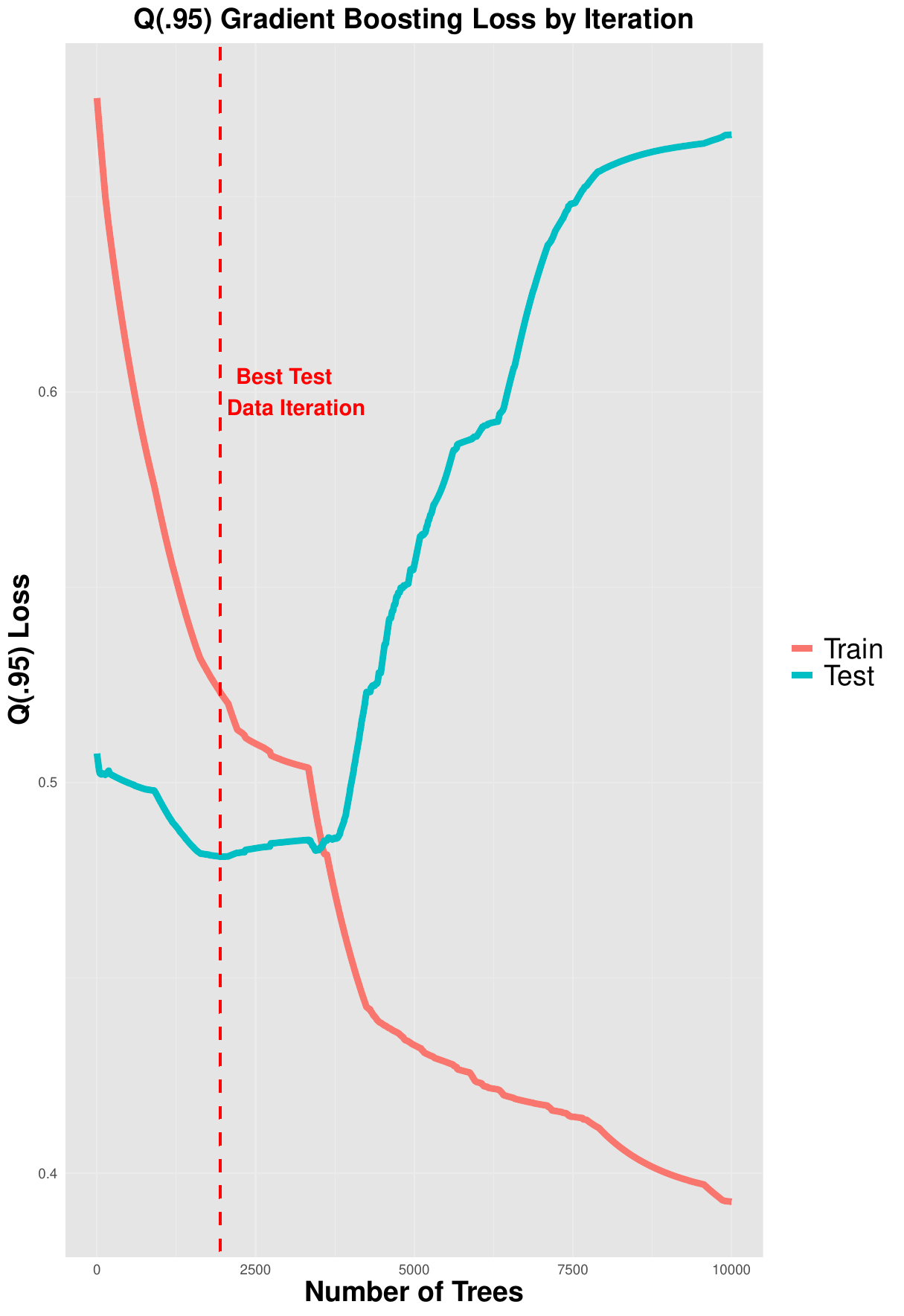}
\caption{Best Test Data Stopping Iteration for the Paris Data Is 1942}
\label{fig:performance}
\end{center}
\end{figure}

The quantile loss function used shortly is defined as:

\[
L_{\tau}(y, \hat{y}) =
\begin{cases}
\tau \cdot (y - \hat{y}) & \text{if } y \geq \hat{y} \\
(1 - \tau) \cdot (\hat{y} - y) & \text{if } y < \hat{y},
\end{cases}
\]
where $y$ is the observed response, $\hat{y}$ is the fitted response, and $\tau$ is the desired quantile. We apply quantile gradient boosting to estimate the 95th quantiles of the conditional temperatures (i.e., $\tau$ = Q(.95)). Quantile loss is designed to be asymmetric such that underestimates can be more costly than overestimates. Using values of $\tau$ larger than .50, makes the gradient boosting algorithm work harder to fit larger values of the response.\footnote
{
The need for asymmetric loss functions has been recognized by leading scholars for decades (Granger, 1969).
} 

All seven predictors were used in training. 10,000 trees were fit sequentially with a shrinkage (``learning rate'') of .001, and an interaction depth of 4. Each additional tree can be seen as an iteration. The intent was to proceed gradually across iterations so that smaller, subtle relationships could be exploited. 

As shown in Figure~\ref{fig:performance}, the fitting process looks routine. The training data fit continually improves while at some point the test data fit begins to degrade. The best test data fit occurs at iteration 1492, which provides a sensible stopping criterion. Figure~\ref{fig:performance} indicates that very similar results follow when as many as 4000 trees are used, which makes the fit relatively robust to the number of iterations employed. There is a hint of double descent where the performance losses for the training data and test data cross, but even if real, the transient drop is not the test data performance minimum. All subsequent results for 2019 derive from the 2019 training data fitted with 1942 iterations/trees. 

Figure~\ref{fig:performance} also implies that the algorithm trained on data from 2019, has some fitting skill with the 2018 data. There are useful similarities even though lagged predictors are measured in the subsequent year.  The earlier assumption that the same physics applies in adjacent years may have some practical merit. Should there be comparable similarities for the 2020 data, useful 2020 forecasts might be obtained from the algorithm trained in 2019.

\begin{figure}[htbp]
\begin{center}
\includegraphics[width=4in]{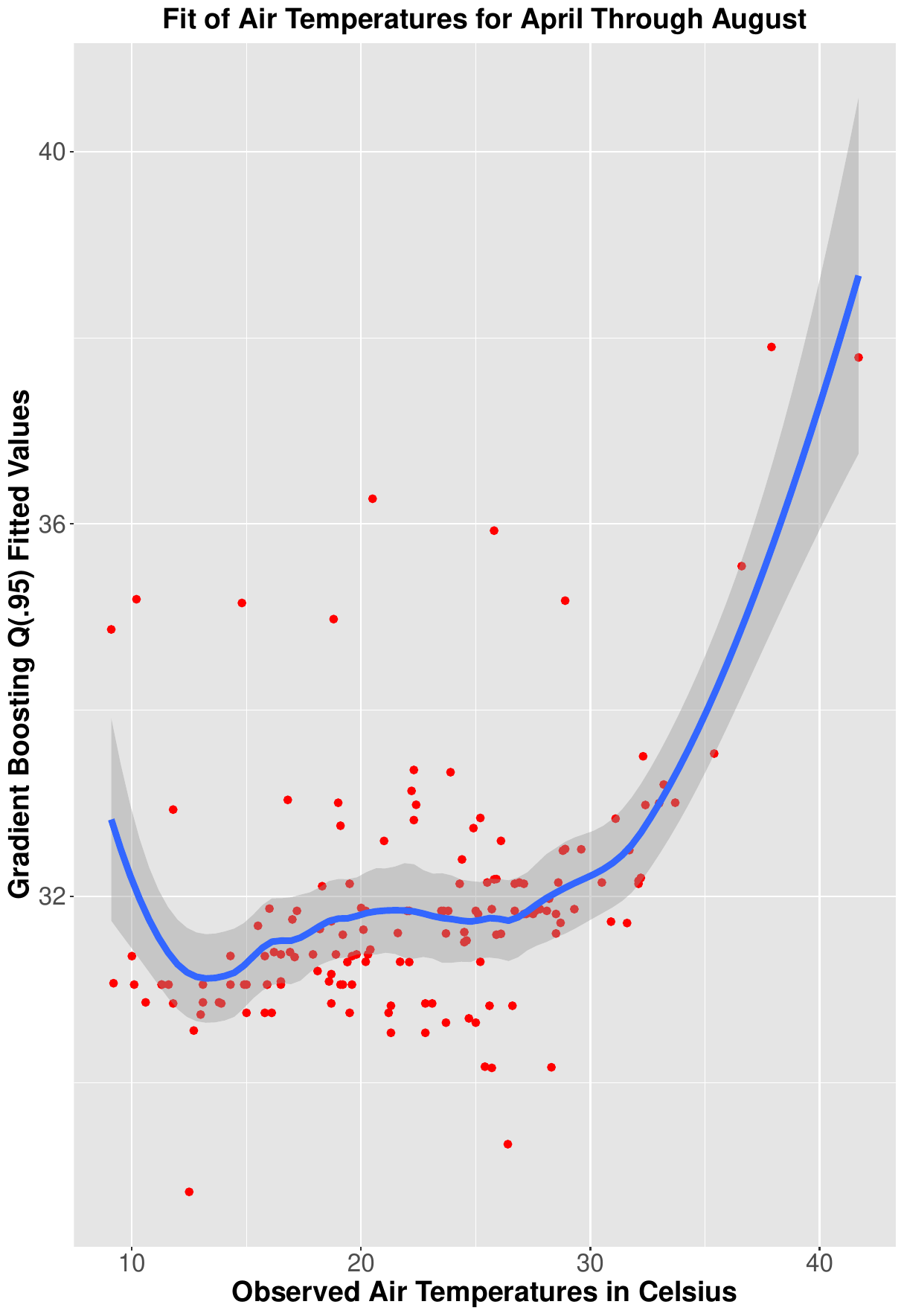}
\caption{Fitted Q(.95) values from quantile gradient boosting in red are plotted against the observed 2pm Paris weather station temperatures. A Loess smoother in blue and conventional 2 SE error bands in gray are overlaid. (N=153)}
\label{fig:fit2019}
\end{center}
\end{figure}

Figure~\ref{fig:fit2019} evaluates how the fitted values for the 95th percentile correspond to the weather station 2pm observed temperatures. Fitted values are in red, a Loess smother is in blue, and an error band is in gray. Except for a few relatively low 2pm temperatures, the general trend is positive and nonlinear. When the observed temperature is warmer, the Q(.95) value, \emph{estimated from predictor values two week earlier}, is higher. The fitted relationship is strongest for the few very warmest days. There are several high fitted values for observed temperatures from 10 to 30\textdegree{}C that might qualify as false alarms.\footnote
{
 The sparsity of extreme high temperatures is not an artifact. Such temperatures empirically are very rare. 
}

 Comparing the Y-axis against the X-axis in Figure~\ref{fig:fit2019} shows that the Q(.95) values are on average substantially larger than the observed values. For example, a fitted values of 32\textdegree{}C corresponds to an observed value of 20\textdegree{}C
. However, with higher observed temperatures, the gap closes, and the very high Q(.95) values tend to be underestimates. Nevertheless, a substantial positive relationship materializes for observed values greater than 30\textdegree{}C with observed values of about 104\textdegree{}F corresponding to fitted values of about 100\textdegree{}F
.\footnote
 {
 The error band is conventional but depends here on several unreasonable assumptions. For example, the data are assumed to be iid, and as a multiple time series, generally they are not.
 }
 
\subsubsection{Predictor Influence}

\begin{figure}[htbp]
\begin{center}
\includegraphics[width=3.5in]{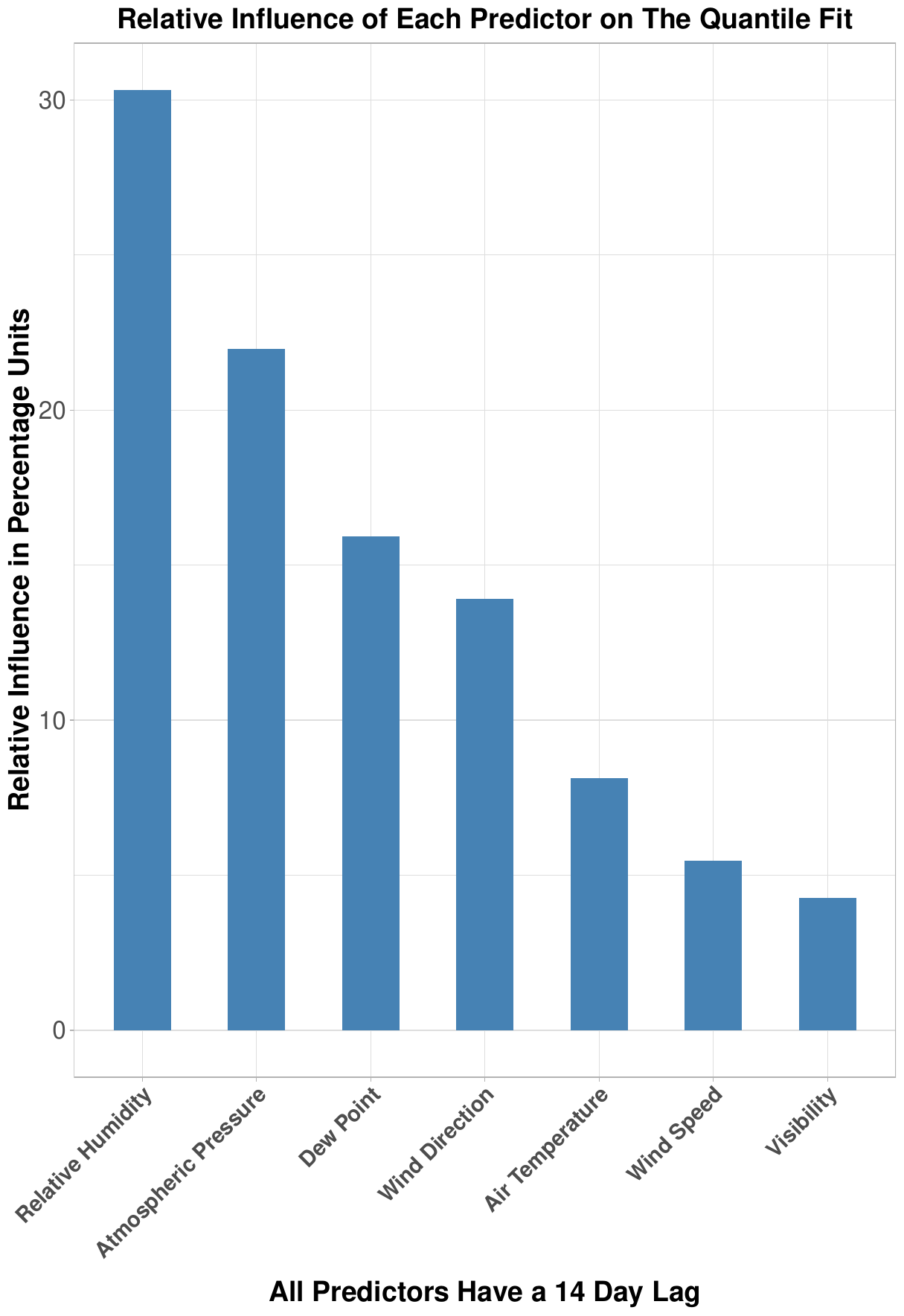}
\caption{Shown is the percentage contribution to the quantile fit for each predictor. Before percentages are computed, each predictor's contribution is the average reduction in quantile loss over boosted trees when that variable is used for splitting.}
\label{fig:influence}
\end{center}
\end{figure}

An algorithm is not a model and, hence, not a proper vehicle for doing causal inference. Still, there can be some interest is documenting the relative importance of each predictor to the overall fit. Figure~\ref{fig:influence} is a bar plot of the relative contribution of each predictor to improvement in the Q(.95) loss (Friedman, 2001: 1217), standardized as percentage contributions. 

More formally, for a collection of $M$ decision trees $\{T_m\}_1^m$ obtained from boosting, an average reduction in estimated loss $\hat{J}^2_j$ over $M$ trees from splits on predictor $j$ is
  \[
\hat{J}_{j}^{2} = \frac{1}{M} \sum_{m=1}^M \hat{I}^2_j (T_m),
\]
where $i_j$ is the reduction in loss (here, quantile loss) for tree $T_m$ from variable $j$. For instance, in Figure~\ref{fig:influence}, relative humidity with a two week lag accounts for about 30\% of the overall relative reduction in quantile loss after the 1942 trees that are grown and aggregated, but all of the predictor variables contribute to the fit. Essentially, contributions to the fit are measures of partial association, not estimated causal parameters.\footnote
{
The relative contributions to the fit add to 100\%.
} 

\subsubsection{Predictor Partial Association With the Response}

Partial dependence plots show for a given predictor the relationship estimated by gradient boosting (or any of several other supervised learning algorithms) for the association between that predictor and the response, with all other predictors held constant at their means (Friedman, 2001: section 8.2).\footnote
{
Partial dependence plots can be generalized to show using a surface (not a line) the joint relationship between a pair pf predictors and the response, all other predictors held constant.
}

More formally, the partial dependence function for a set of features \( S \subset \{1, \dots, p\} \) is defined as:

\[
f_S(x_S) = \mathbb{E}_{X_C}\left[ f(x_S, X_C) \right] = \int f(x_S, x_C) \, dP(x_C)
\]

where:
\begin{itemize}
    \item \( x_S \) is the feature of interest (e.g., relative humidity),
    \item \( C = \{1, \dots, p\} \setminus S \), the set of indices of all features excluding those in \( S \) (i.e., the complementary features),
    \item \( f(x) \) is the fitted function (e.g., from gradient boosting),x
    \item \( P(x_C) \) is the joint distribution of the complementary predictors.
\end{itemize}

In practice, this is approximated by averaging over the empirical distribution of the training data:

\[
\hat{f}_S(x_S) = \frac{1}{n} \sum_{i=1}^n f(x_S, x_{C}^{(i)})
\]
where \( x_C^{(i)} \) is the \( i \)-th observation of the complementary predictors in the training data. To speed computation, the data are often binned. The specific way in which 
partial dependence is constructed depends on the learning algorithm.

\begin{figure}[htbp]
\begin{center}
\includegraphics[width=3in]{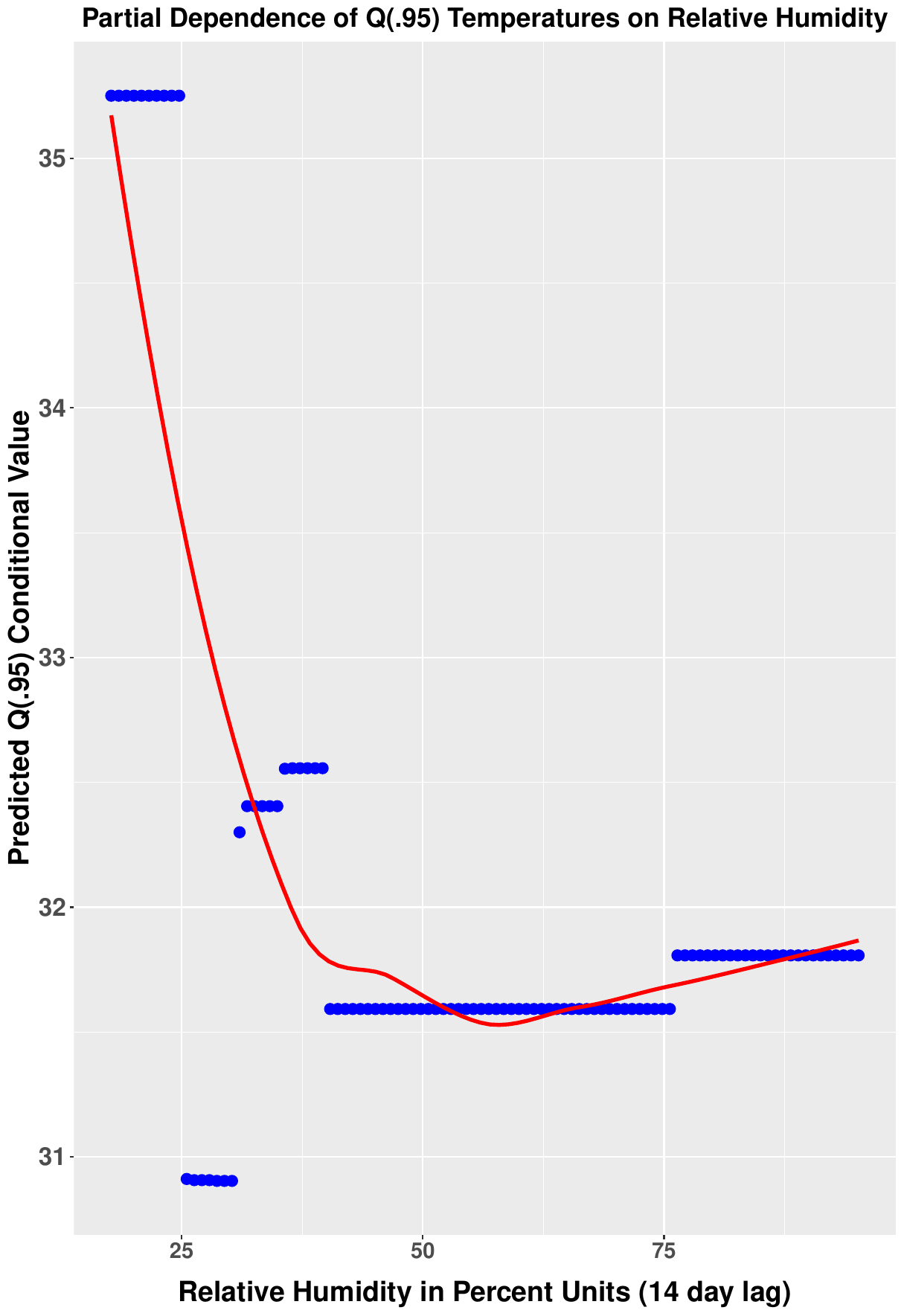}
\caption{A partial dependence plot of the relationship between relative humidity and the conditional Q(.95) temperature quantiles, with all of the other predictors fixed at their means.}
\label{fig:partial}
\end{center}
\end{figure}

There are seven such partial dependence plots from this analysis. All are substantially nonlinear. To illustrate, Figure~\ref{fig:partial} shows how the predictor relative humidity is related to Q(.95) celsius temperatures with all other predictors ``held constant." One has the average the Q(.95) temperature quantile for each reported value of relative humidity with all other predictors fixed at their means. Lower humidity is associated with higher temperatures two weeks later, but the movement from higher to lower temperatures appears to proceed in steps. This may result from rounded Q(.95) temperature values. Just as with contributions to fit, Figure~\ref{fig:partial} says little about causality. Nevertheless, Figure~\ref{fig:partial} implies that nonlinearity would likely be an important challenge subsequently should causal analyses be undertaken.

\subsubsection{Forecasting Performance}

\begin{figure}[htbp]
\begin{center}
\includegraphics[width=4in]{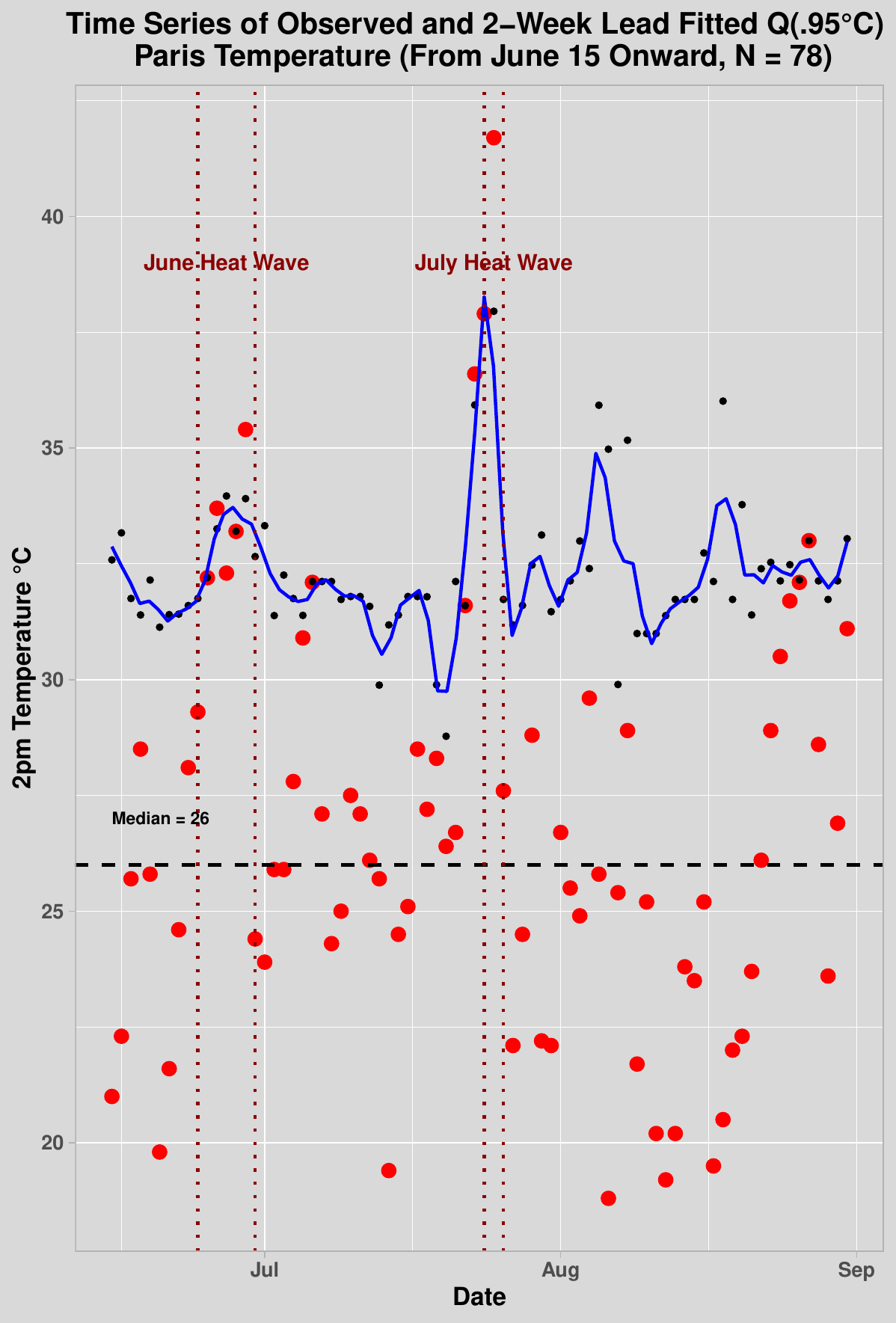}
\caption{The gradient boosting fit in a time series format. The red dots are the observed temperature values, the black dots are the Q(.95) fitted values produced from predictor values two week earlier, and the blue overlay is a loess smoother applied to the fitted values and serves as a visual aid. The vertical dotted lines show the dates on which an extreme heat wave was reported, which was considered \emph{after} the figure was initially constructed.}
\label{fig:TS2019}
\end{center}
\end{figure}

We have now arrived our Paris trip's 2019 destination: actual forecasting. Figure~\ref{fig:TS2019} formats and plots the gradient boosting fitted values constructed earlier as a time series, along with the observed air temperature values, and a loess smoother serving as a visual aide overlay on the fitted values. Because the fitted values target the 95th percentile, they generally fall above the observed temperatures. All are a product of predictor values two weeks earlier.\footnote
{
Figure~\ref{fig:TS2019}, only displays the data after June 15th even though the full dataset was used to in the analyses. Including the data from March and the first two weeks in April makes the figure far more cluttered and provides no additional insights.
}

There are two spikes in the smoother that correspond to substantial increases is the observed air temperatures. The spikes appear for the last week in June and third week in July. Smaller but still noticeable spikes in August are a product of the fitted values alone and might be interpreted as ``false alarms.'' 

After the two large spike were observed, a search was undertaken for information on heat waves in Paris during the summer of 2019. Wikipedia provides a list of heat waves from around the world including for the summer of 2019 \url{(https://en.wikipedia.org/wiki/List_of_heat_waves#2019)}. The spikes in June and July correspond to the wikipedia reports of heat waves. The BBC \url{(https://www.bbc.com/news/world-europe-49628275)} and the Weather Channel provide more details, including that nearly 1,500 people died from the heat (\url{https://weather.com/news/news/2019-09-08-summer-heat-waves-kill-1500-people-in-france}). Our analyses show that these two heat waves can be anticipated by about two weeks. 

Under two conditions, conformal prediction regions can be used to address forecasting uncertainty: (1) The data and the trained algorithmic structure are treated as given and (2) the residuals from the training are exchangeable).\footnote
{
``It might be surprising to learn that, with conformal prediction, rigorous uncertainty quantification is
possible without any assumptions at all on the model, and with minimal assumptions on the distribution of the data—no asymptotics, no limit theorems, and no Gaussian approximations'' (Angelopoulos et al., 2025: page 4).
}
Even for multiple time series data, the residuals sometimes have some lingering temporal structure. From the gradient boosting discussed above, the residuals evidence a modest autoregressive structure implying weak dependence. Our conformal uncertainty estimates would then only apply asymptotically. Consequently, we applied an AR(1) time series model (Hyndman and Athanasopoulos (2021: section 9.2)) to the residuals producing a coefficient of .42. The residuals of that AR(1) time series model were indistinguishable from white noise. Those are the residuals used as nonconformal scores going forward, arguably because exchangeabilty becomes plausible. We proceeded with an extensions of the work of Romano and colleagues (2019) by using quantile gradient boosting to estimate the lower and upper quantile bounds. See also Gibbs and Cand\`es.

To ground the Q(.95) forecasting discussion, we drew a random sample of 100 days from the late spring and summer of 2020, and provisionally assume that an algorithm trained on Paris data from 2019 can provide useful Paris forecasts for 2020; the underlying physics does not change. In these data, the observed temperatures are known. That would not be true in real applications; if the temperatures were known, there would be no need to forecast them. For this evaluation, knowing the ``truth'' helps to convey what conformal prediction regions mean. 

2020 observed temperatures range from 10.2 to 36.4\textdegree{}C. 2019 forecasts range from 28.1 to 37.5\textdegree{}C. The mean for the observed temperatures is 22.7\textdegree{}C. The mean for the forecasted Q(.95) temperature is 32.1\textdegree{}C. The summer of 2020 did not have as many very extreme high temperatures as 2019 and had a larger fraction of relatively cool summer days. Still, the results do not seem to contradict the forecasting approach applied. Daily 2pm Paris temperatures are realizations of very noisy random variables.

How good are the forecasts? We continue to assume that the data from 2020 are realized from the same underlying physics as for 2018 and 2019 and that, therefore, forecasts two weeks in advance can capitalize on the same physical relationships. Stipulated is a conformal coverage probability of .70; the actual coverage probability will be \emph{at least} .70. There is nothing special about that figure, which can be appropriate in some practical settings. The knee-jerk stipulated conformal coverage probability is usually .95.\footnote
{
A larger stipulated probability implies greater confidence that the actual outcomes fall within the conformal prediction regions. But the necessary price of less precision; the conformal prediction interval is wider. Such tradeoffs are subject-matter decisions, not statistical decisions, that ideally should be made before the forecasting analysis begins.
}

The Q(.95) 2020 forecasts come with on the average a conformal interval of about $\pm$ 2\textdegree{}C, ranging from a minimum of approximately $\pm$ 1.5\textdegree{}C
 to a maximum of approximately $\pm$ 3\textdegree{}C.\footnote
{
Each new observation can have its own confidence region with its own precision. In that sense, the conformal regions are ``adaptive'' (Romano et al., 2019).
}
Of the 100 sampled days from 2020, 12 had observed temperatures falling within their Q(.95) prediction region. The majority of actual temperature fell below their lower prediction region bounds. That makes sense because the forecasts are for the 95th percentile. If one considers only the 10 highest observed temperatures, all 10 fall in their conformal prediction regions. Forecasts near or above Q(.95) temperatures work well two weeks in advance. In this instance, the coverage probability might be a bit larger than .70.\footnote
{
Given the coverage probability of at least .70, a little less than 4 of the 12 successes are likely to be false successes. Of the remaining successes, all are more likely to have the real outcome 2 weeks later within their prediction regions. It may be possible in the future to improve the precision of the prediction regions building on extreme value distributions, but additional assumptions seem to be required to look beyond the empirical tails of the conformal distribution (Velthoen et al., 2023; Pasche et al., 2025). Temporal dependence is a significant complication.
}
 
\begin{figure}[htbp]
\begin{center}
\includegraphics[width=4in]{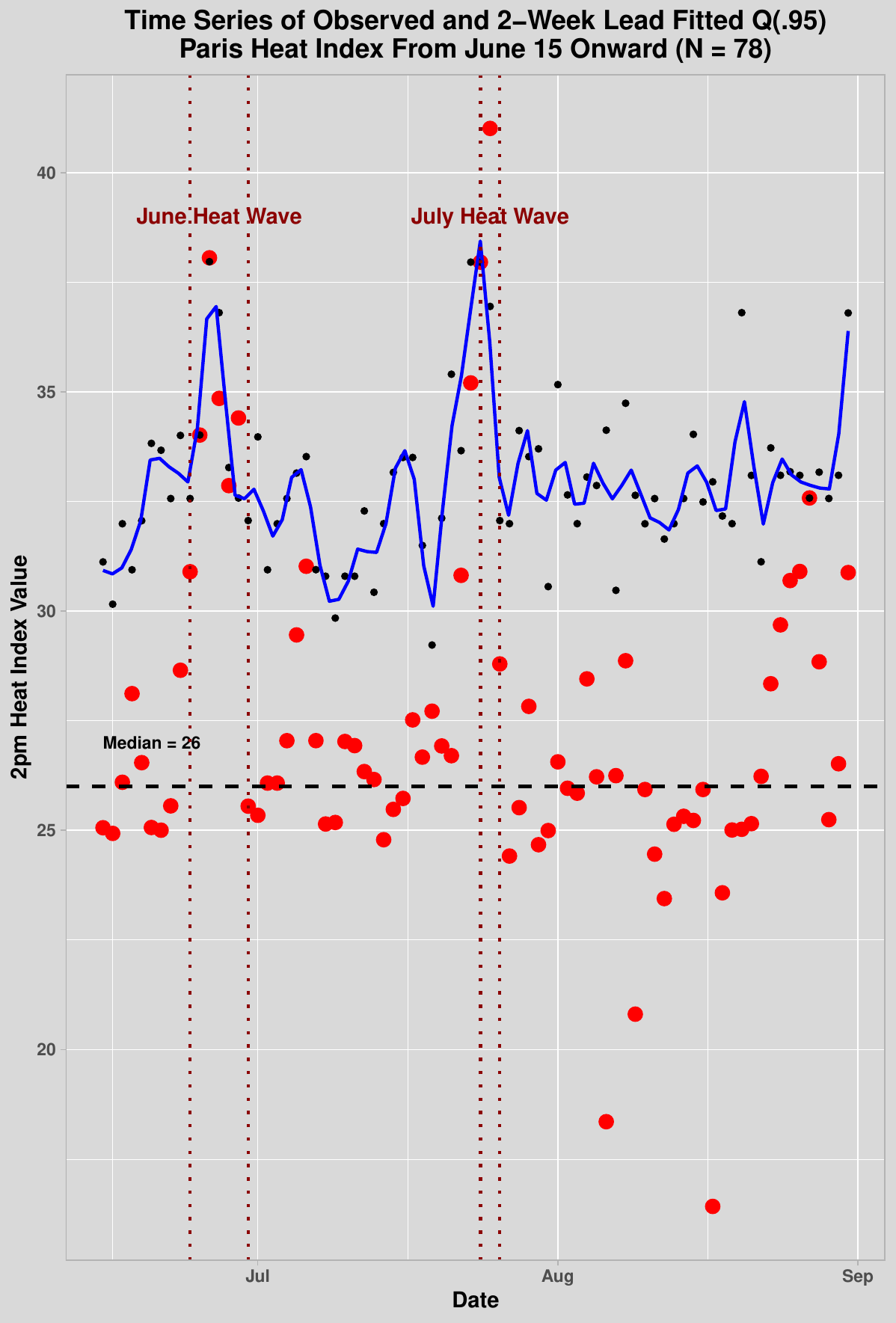}
\caption{The gradient boosting fit for the Steadman heat index in a time series format. The red dots are the observed heat index values, the black dots are the Q(.95) fitted heat index values produced from predictor values two weeks earlier, and the blue overlay is a loess smoother applied to the fitted values and serves as a visual aid. The vertical dotted lines show the dates on which an extreme heat wave was reported, which was considered \emph{after} the figure initially was constructed.}
\label{fig:heatindex}
\end{center}
\end{figure}

The full analysis above was repeated with the Steadman Heat Index computed using humidity, not dew point. The humidity enhanced heat index heterogeneity for the fitted values,  especially making high temperatures substantially larger. Nevertheless, all of the new results are very similar to the results discussed above. In particular, Figure~\ref{fig:heatindex} is much like Figure~\ref{fig:TS2019} but with the very high temperatures inflated somewhat. The main consequence is that the two spikes become more pronounced relative to the many smaller spikes. But the smoother is just a convenience. It is important to inspect where the red dot and black dot fall on a given day for both Figure~\ref{fig:heatindex} and Figure~\ref{fig:TS2019}. For the higher observed temperatures, the two ideally are very close to one another. When they are not, the lagged predictors and gradient boosting algorithm likely are to blame. But measurement error and noisy data realization cannot be ruled out. 

Analyses of the two other pairs of years was comparable. All of the statistical procedures used for 2018 and 2019 performed about the same for 2020 and 2021, and for 2022 and 2023. It was visually evident that when a few days with very high temperatures were forecasted, they corresponded quite well to dates of reported heat waves. The main technical challenge was that for some Steadman heat index analyses, the \textit{gbm} procedure failed to iterate after the first fit was completed. A likely cause is a very small number of days with unusually extreme Steadman heat index values. Once they are fit quite well after a single iteration, other variables cannot improve the fit. For the additional two pairs of years, plots similar to Figure~\ref{fig:heatindex} and Figure~\ref{fig:TS2019} are easily produced if requested.\footnote
{
The Steadman heat index also was computed using dew point instead of relative humidity Steadman, 1984). In these data, The two correlate above .99. There was no need to repeat the heat index analyses with the Steadman dew point measure.
}

\section{Results for Cairo}

The decision to forecast extreme and rare temperatures for Cairo was an attempt to push the envelope. Even working with data from the same kinds of sources organized in the same manner, the forecasting task would be more difficult. Paris abuts the Loire Valley. Cairo abuts the Eastern Desert; the Sahara Desert east of the Nile River and west of the Red Sea.  Desert climates can experience very rapid and dramatic changes in temperature depending literally on which way the wind blows.  Wind direction also can change rapidly and dramatically. There can be for Cairo relatively cool temperatures when winds from the Mediterranean blow in from the north that are quickly overwhelmed when a high barometric pressure over the Red Sea or Arabian Peninsula combines with a low pressure trough to the east (e.g., over Libya). The Khamsin winds blowing in from the Eastern Desert typically bring hot dry air that can quickly raise air temperatures by more than 10\textdegree{}C
 and temperatures that spike to 40\textdegree{}C or more. But the confluence of such events can be very short-lived, and their temperature effects can depend heavily on a tipping point in wind directions and velocities. Very shortly after one wind direction overcomes another wind direction, changes in temperature quickly might materialize.\footnote
{
This process is experienced along the southern California coast when an onshore wind from the Pacific ocean, which brings cloud cover, dampness and low temperatures, is overcome by offshore winds often originating in the Great Basin and upper Mojave desert. (https://en.wikipedia.org/wiki/Santa\_Ana\_winds). A transition to clear skies, low humidity, and substantial temperature increases can materialize within an hour or less.
}
Warner (2004) offers a rich account of desert climates that is well beyond the scope of this paper.

\begin{figure}[htbp]
\begin{center}
\includegraphics[width=4in]{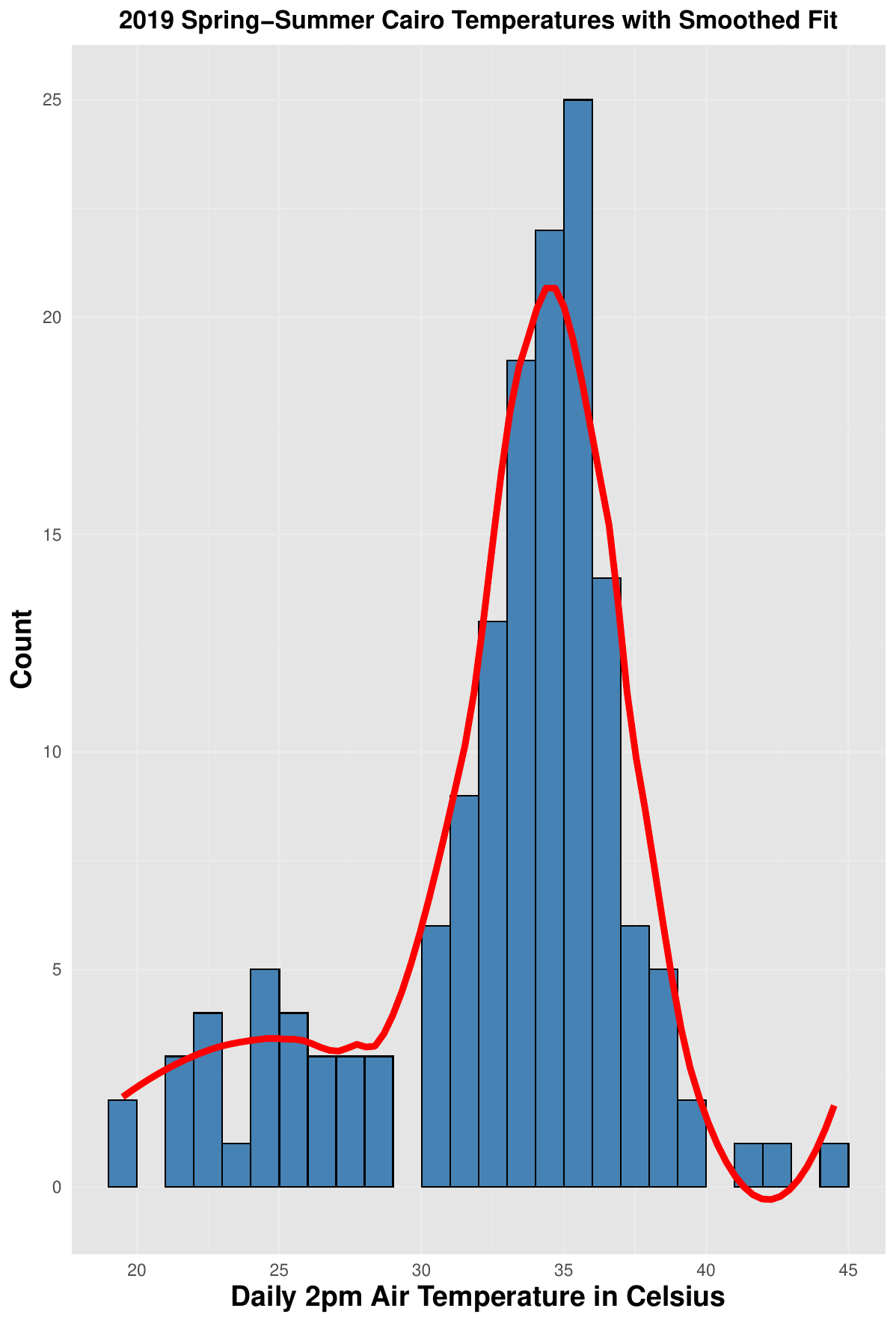}
\caption{2pm Cairo Air Temperatures March Through August 2019 with a Density Smoother Overlay}
\label{fig:hist2019C}
\end{center}
\end{figure}

The data come from the weather station at the Cairo International airport, which is located in the suburb of Heliopolis about 15 miles northeast of Cairo. Temperatures at the airport are often a bit cooler than temperatures in downtown Cairo. But temperatures in the two locations tend to move together. As before, air temperature at 2pm is the response variable. The predictors are all lagged by two weeks. Atmospheric pressure is of no use because all of its values are missing. In retrospect, that could be a substantial loss as a precursor of Khamsin winds. Air pressure readings from strategically  positioned desert locations to the west and to the east might be even more helpful.

Figure~\ref{fig:hist2019C} shows histogram of the air temperatures from March through August, just as earlier. But the shape is very different. Efforts to fit a GEV function as an overlay failed. The smother in red is a conventional Loess density smoother. There are now tails to the left as well as the right. Extreme high temperatures at quite rare and might even be considered outliers because of the spaces between their bins and the mass of the data. 

\begin{figure}[htbp]
\begin{center}
\includegraphics[width=4in]{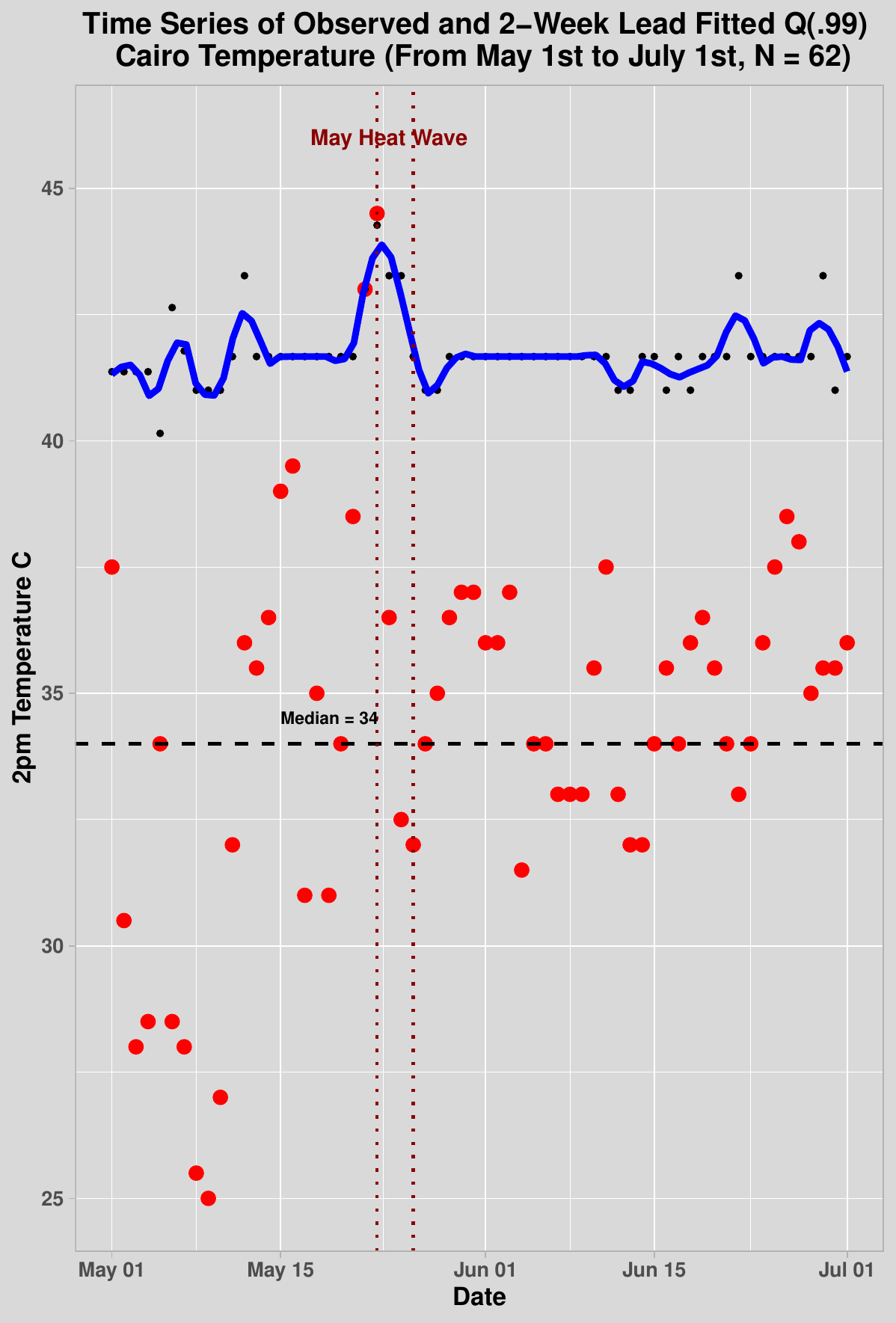}
\caption{Cairo gradient boosting fit in a time series format. The red dots are the observed temperature values, the black dots are the Q(.99) fitted values produced from predictor values two week earlier, and the blue overlay is a loess smoother applied to the fitted values and serves as a visual aid. The vertical dotted lines show the dates on which an extreme heat wave was reported, which was considered \emph{after} the figure was initially constructed.}
\label{fig:ts2019C}
\end{center}
\end{figure}

Figure~\ref{fig:ts2019C} is a format used earlier, but now the fitted values for Cairo temperatures in a time series layout. Gradient boosting was used as before, but it had to work harder to fit the high rare values; Q(.99) rather than Q(.95) was the target quantile specified. Still the fit overall was not very good. For example, we had wind direction and speed two weeks earlier. Having a lag of a day or two instead might have helped a great deal, but that would not have allowed for much advanced warning. 

\begin{figure}[htbp]
\begin{center}
\includegraphics[width=4in]{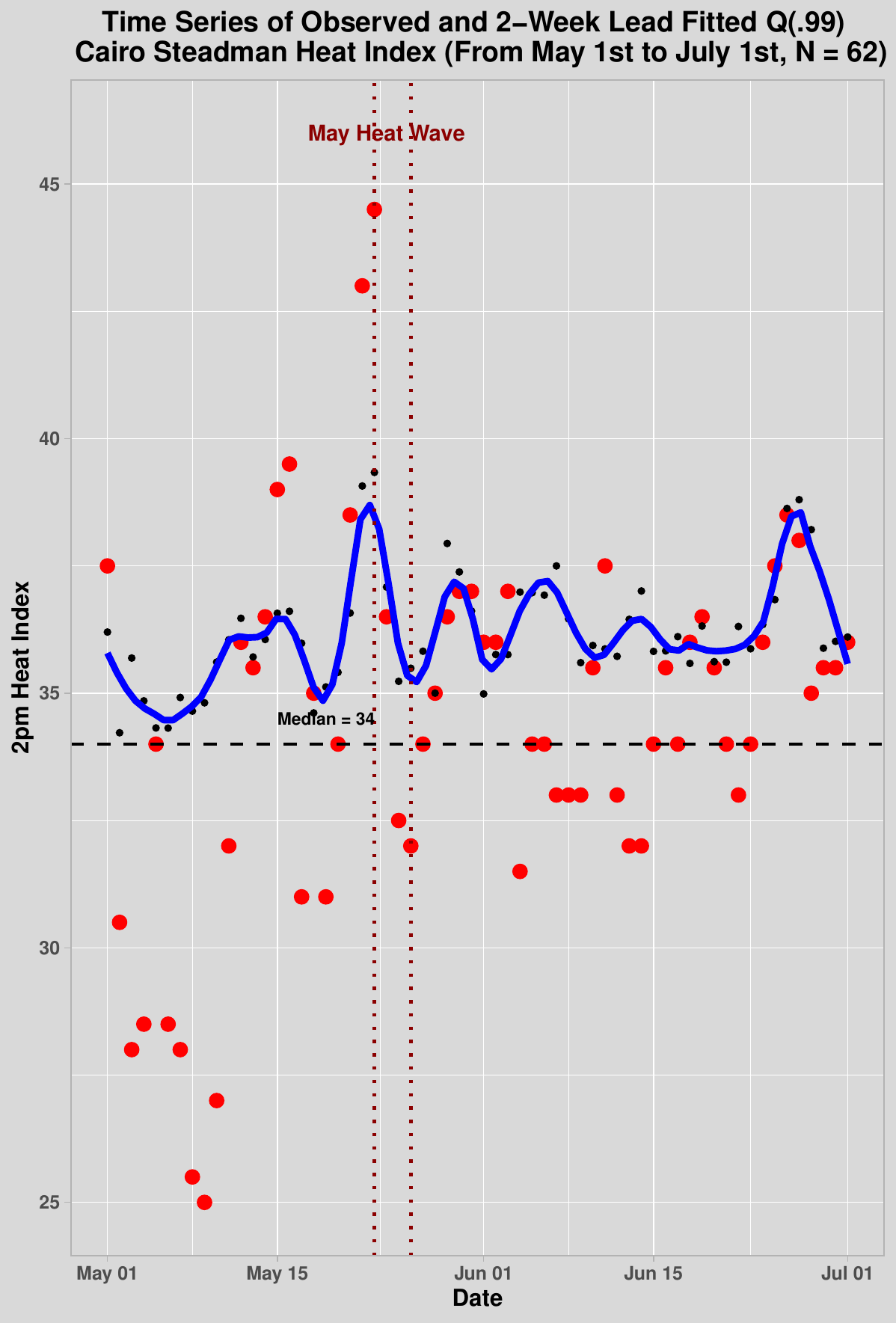}
\caption{Cairo gradient boosting fit in a time series format. All of the data were used for the gradient boosting, but some of data at either end were not plotted to help reduce clutter. The red dots are the observed Steadman heat index values, the black dots are the Q(.99) fitted values produced from predictor values two week earlier, and the blue overlay is a loess smoother applied to the fitted values and serves as a visual aid. The vertical dotted lines show the dates on which an extreme heat wave was reported, which was considered \emph{after} the figure was initially constructed.}
\label{fig:heatindex2019C}
\end{center}
\end{figure}

Still, Figure~\ref{fig:ts2019C} provides some interesting results. There is a prominent spike predicted two weeks in advance that we later learned corresponded quite well to one very high and one record observed temperature that some called a brief May heat wave appearing unusually early.  The rest of the temperature data are fit rather poorly.

Finally, Figure~\ref{fig:heatindex2019C} repeats the analysis and plotting using the Steadman heat index. Humidity seems to play a small role in how Cairo temperatures are experienced, but it shuffles around the observed data a bit. The pair of extremely high May values is again captured, but less well with the observed index values substantially underestimated. There are also several other noticeable fitted peaks. As a result, the overall correspondence between the fitted index values and the observed index values is much improved. But there is no longer a clear narrative.

In summary, the weaker heat index results still provide for a reasonable warning two weeks in advance of some high index days. Given the importance of wind direct and wind speed, more success might follow from substantially shortened the lag lengths. But the warning between the forecast and the events forecasted then could be only a day or two apart. The Steadman heat index made the results far less definitive, perhaps because humidity might be a relatively minor factor in Cairo weather. 

\section{Implications for Policy}

For both Paris and Cairo, the same physics is assumed to apply year to year. But thermodynamics and other related processes can create large variation in the observed random variables over time. Consequently, the summer of 2019 may have very different weather from the summer of 2020. An important result is that an empirical analysis from an earlier year may lead to prediction regions that are on the average quite different from predictions regions from a later year. An important result is that the Q(.95) values for one year may be a misleading benchmark for another year. 

The consequences may differ too. For example, many people may die from Q(.95) temperatures one year and none may die from Q(.95) temperatures another year. There can be comparable complexity by geographical location because some locales have better infrastructure for adapting to hot weather. Ecosystem vulnerability can vary widely too; compare Seattle, Washington to Phoenix, Arizona. These and related concerns suggest that each local setting should determine its own threshold for the level at which temperatures are unusually threatening to public health and local ecosystems. 95\textdegree{}F may be a major threat for the Russian far east but not for the Mojave desert in the southwestern United states (Berk et al., 2025).

Such framing suggests the follow actionable steps linking forecasts of rare, extreme, hot temperatures to their consequences that can matter.
 \begin{enumerate}
 \item
 Train quantile \textit{gbm} on high quantile temperatures at a common, instructive solar time each day for an informative set of late spring and summer months. By definition, these temperature must be least relatively very rare and higher than typical. The summer months might be chosen from the year immediately preceding the year for which heat forecasts are desired. Predictors that may well differ depending on the data, should be lagged with the number of days determined by policy considerations and requisite forecasting accuracy. Test data from another year is likely needed.
 \item
 A risk threshold high temperature should be locally determined based on public health and ecological impact (e.g., 35\textdegree{}C).
  \item
 When forecasts are needed, prediction regions that are too low to include the threshold temperature can be ignored.  Prediction regions that include the threshold high temperature should be taken seriously.\footnote
 {
 Each prediction region is located by the day's forecasted temperature. It might make sense to update the trained algorithm's results as new data in the selected year needing forecasts became available. But updating the training data is beyond the scope of this paper.
 }
\end{enumerate} 

In short, forecasted prediction regions would as a policy matter be concerning if they contain the local risk threshold. But actionable forecasts require high quality, timely data.
\footnote
{
The forecast would be contained as well because it anchors the prediction region.
} 

Finally, there seems little hope that one trained algorithm will work effectively in many different settings. So far at least, transfer learning has not been productive either. Perhaps heat forecasting algorithms need to trained on local data from where the forecasts are likely to be used. The Paris to Cairo differences are one telling illustration.

\section{Discussion}

The forecasting results are not cherry picked. Widespread European and Middle East heat waves reported over the past several summers provided no more than motivation. Paris in particular received extensive media coverage, has a healthy respect for science and a well-respected science infrastructure. The Paris-Montsouris weather station was a natural and the first stop. The sequence of years was selected to capture the recent impact of global warming. Cairo was picked as a more  challenging setting.

A major concern is whether the key forecasting assumption of stable climate physics year to year holds in sufficient detail. For development of forecasting tools, the approach presented uses late spring and all summer months from one year for training data, late spring and all summer months from an adjacent year for test data, and late spring and all summer months from another adjacent year for evaluating forecasting skill. It seems that in practice the data do not have to be a literal equivalent of random realizations from the exact same set of physical processes over adjacent years, but the closer the approximation the better. All of the daily observation stem from random variables, which complicates empirical comparisons across years. More work is needed on how to judge setting comparability, especially with global warming perhaps accelerating and disruptive climate events such as El Ni\~no. 

Nevertheless, It may be that similar forecasting procedures could be easily developed in many other locations because weather stations commonly retain and share data. There are about 14,000 active weather stations in the current NOAA ISD collection. Some have more extensive data than others, but the Paris and Cairo analyses could be easily repeated in many locations around the world. Currently, the data can be obtained at no cost and is easily downloaded. 

As the same time, there are some important limitations. Perhaps most important, many of the key variables of interest to climate scientists are not available from weather stations. If the scientific goal is explanation, weather station data will probably not help much. For forecasting, however, there may be real promise. Moreover, the promise may be quite general. Berk and colleagues (2024; 2025) working with AIRS data also report some success trying to anticipate extreme hot and rare weather events two weeks in advance. AIRS data are produced through remote sensing by a satellite designed for world wide coverage. That satellite will soon be decommissioned, but there is hope and some plans for an upgraded replacement \url{(https://nisar.jpl.nasa.gov/?utm_source=chatgpt.com).} Other data-gathering satellites are anticipated. So far at least, however, their spatial and temporal scales are likely to be much coarser that what data weather stations provide. The popular option treating model simulation output as data (e.g., from the Community Earth System Model, CESM) seems designed to address major trends over time and not extreme, rare phenomena. In the future, smart downscaling might help, but the difficulties are substantial (Gettleman and Rood (2016). 

There are also matters of timeliness. If forecasting is the goal, requisite data must be relatively easy to obtain within two weeks of the rare climate event being forecasted. This implies ongoing measurement rapidly packaged for internet distribution. The World Meteorological Organization (WM0) is a good source of information about available data, although direct access to data is provided through WMO-partnered centers or national meteorological services. Useful organizations to contact include the National Oceanic and Atmospheric Administration (NOAA), the European Centre for Medium-Range Weather Forecasts (ECMWF), and the European Organization for the Exploitation of Meteorological Satellites (EUMESAT). In the medium term at least, NOAA is likely facing substantially reduced resources. One might start with ECMWF and EUMESAT. 

Whatever the data sources, forecasts derived from machine learning will be a major, and even dominant, statistical tool (Pathak et al., 2022; Price et al., 2024; Bodnar et a., 2025). Advances in estimating forecasting uncertainty using approaches with lean assumptions will also contribute mightily. The statistical methods used in this paper are but an early example. 

\section{Conclusions}

The foresting results for Paris suggest that instructive forecasts of unusually high temperatures can be obtained at least two weeks in advance from weather station data. The case is weaker for Cairo. Still, such data are easily obtained at no cost in a timely manner. An effective data analysis can follow from machine learning and conformal prediction regions. However, the overriding goal is forecasting. Prospects from weather station data for explanation that advances the scientific discourse are at best modest. 

\section*{References}
\begin{description}
\item
Ahmed, S., Nielsen, I.E., Tripathi, A. et al. (2023) ``Transformers in Time-Series Analysis: A Tutorial.'' \textit{Circuits, Systems, and Signal Processing} 42: 7433 -- 7466.
\item Belkin, M., Hsu, D.H., and Mandal, S. (2019) ``Reconciling Modern Machine-Learning Practice and the Classical Bias–Variance Trade-Off.'' \textit{Proceedings of the National Academy of Sciences} 116(32): 15849 -- 15854.
\item
Angelopoulos, A.N., Barber, R.F., and Bates, S. (2025) ``Theoretical Foundations of Conformal Prediction.'' arXiv2411.11824v2 [mathST].
\item
Berk, R.A., Braverman, A., and Kuchibhotla, A.K. (2024) ``Algorithmic Forecasting of Extreme Heat Waves.'' arXiv:2409.18305v4 [stat.AP].
\item
Berk, R. and Braverman, A (2025) ``Forecasting Extreme Temperatures in Siberia Using Supervised Learning and Conformal Prediction Regions.'' arXiv:2503.16118v1 [stat.AP].
\item
Bodnar, C., Bruinsma, W.P., Lucic, A. et al. (2025) ``A Foundation Model for the Earth System.'' \textit{Nature}  publisjed online https://www.nature.com/articles/s41586-025-09005-y\#citeas
 \item
 Chernozhukov, V., W\"uthrich, K., and Zhu, Y. (2018) ``Exact and Robust Conformal Inference Methods for Predictive Machine Learning with Dependent Data.'' \textit{Proceedings of Machine Learning Research} 75: 1 -- 17.
 \item
 Coles, S. (2001) \textit{An Introduction to Statistical Modeling of Extreme Values}. Springer
 \item
 Cvijanovic, V., Mistry, M.N., Begg, J.D., Gasparrini A., and Rod\'o, X. (2023) ``Importance of Humidity for Characterization and Communication of Dangerous Heatwave Conditions.'' \textit{npj Climate and Atmospheric Science} 6(33).
 \item
 Friedman, J.H. (2001) ``Greedy Function Approximation: A Gradient Boosting Machine. ''\textit{TheAnnals of Statistics} 29(5): 1189 -- 1222.
 \item
 Friedman J.H. (2002) ``Stochastic Gradient Boosting.'' \textit{Computational Statistics \& Data Analysis} 38 (4):  367 -- 378.
 \item
 Gettleman, A., and Rood, R.B. (2016) \textit{Demystifying Climate Models: A users Guide to Earth System Models} Springer.
 \item
 Gibbs, I., and Cand\`es, E. (2021) ``Adaptive Conformal Inference Under Distribution Shift.'' in M. Ranzato,  A. Beygelzimer, Y. Dauphin, P.S. Liang and J. Wortman Vaughan, \textit{Advances in Neural Information Processing Systems} 34: 1660 -- 1672.
 \item
Graff, A., (2025) ``101 Degrees in May? Even for Texas, This Is Hot,''  \textit{New York times} May 4th, \texttt{https://www.nytimes.com/2025/05/14/weather/ texas-heat-record-may.html}.
 \item
 Granger, C.W.J. (1969) ``Prediction with a Generalized Cost of Error Function. \textit{Operational Research Quarterly} 20: 199 -- 207.
 \item
He, C., Kim, H., Hashizume, M., Lee, W., Honda, Y., and Kim, S.E. (2022) ``The Effects of Night-Time Warming on Mortality Burden Under Future Climate Change Scenarios: A Modelling Study.'' \textit{The Lancet Planetary Health} 6(8), 648 -- 657.
 \item
 Hyndman, R.J. and Athanasopoulos, G. (2021) \textit{Forecasting: Principles and Practice}, third edition at http/OTexts.c0m/fpp3/.
 \item
 Hopke, J. E. (2019). Connecting Extreme Heat Events to Climate Change: Media Coverage of Heat Waves and Wildfires. \textit{Environmental Communication} 14(4), 492–508. 
 \item
 Hulme, M., Dassai, S., Lorenzoni, I., and Nelson, D.R., (2008) ``Unstable Climates: Exploring the Statistical and Social Constructions of `normal' Climate.'' \textit{Geoforum} 40: 197 -- 206.
 \item
 Jacoby, W.G. (2000) ``Loess: A Nonparametric, Graphical Tool For Depicting Relationships Between Variables.'' \textit{Electoral Studies} 19(4): 577 -- 613.
 \item
 Jaque-Dumas, V., Ragine, F., Borgnat, P., Arbry, P., and Bouchet, F. (2022) ``Deep Learning-Based Extreme Heatwave Forecast'' \textit{Frontiers in Climate} 4 - 2022.
 \item
 Klingh\"ofer, D. Braun, M., Br\"uggmann, D.,  and Groneberg, D.A. (2023) ``Heatwaves: Does Global Research Reflect the Growing Threat in the Light of ClimateChange?'' \textit{Globalization and Health} (19)56: 1 -- 17.
 \item
 Leeper, R. D., Harrington, T., Palecki, M. A., DePolt, K., Scott, E., Runkle, J., and Diamond, H.J. (2025) ``The Influence of Drought on Heat Wave Intensity, Duration, and Exposure.'' \textit{Journal of Applied Meteorology and Climatology} 64: 425–438,
 \item
 Li, X., Mann, M.E., Wehnerb, M.F., Rahmstorf, S., Petric, S., Christiansena, S. and Carrilloa, J." (2024) ``Role of Atmospheric Resonance and Land–Atmosphere Feedbacks as a Precursor to the June 2021 Pacific Northwest Heat Dome event.'' \textit{PNAS: Earth Atmospheric, and Planetary Sciences} 121(4): 1 -- 7.
 \item
 Liu, J-M., Ai, S-Q., Jin-Lei Qi, J-L., Wang, L-J., Zhou, M-G., Wang, C-J., Yin, P., Hua-Liang Lin, H-L. (2021) Defining Region-Specific Heatwave in China Based on a Novel Concept of  ``Avoidable Mortality for Each Temperature Unit Decrease.'' \textit{Advances in Climate Change Research} 12(5): 
Volume 12, Issue 5: 611 -- 618,
 \item
Mann, M.E., Rahmstorf, S., Kornhuber, K., and Steinman, B.A. (2018) ``Projected Changes in Persistent Extreme Summer Weather Events: The Role of Quasi-Resonant Amplification.'' \textit{Science Advances} 4(10) DOI: 10.1126/sciadv.aat3272.
\item
Marx, W., Haunschild, R., and Bornmann, L. (2021) ``Heat Waves: A Hot Topic in Climate Change Research.'' \textit{Theoretical and Applied Climatology} 146: 781 -- 800.
\item
McKinnon, K. and Simpson, I.R. (2022) ``How Unexpected Was the 2021 Pacific
Northwest Heatwave?'' \textit{Geophysical Research Letters} 49(18) (e2022GL100380).
\item
Oliveira, R.I., Ornstein, P., Ramos, T., and Romano, J.V. (2024) ``Split Conformal Prediction and Non-Exchangeable Data.'' \textit{Journal of Machine Learning Research} 25:1-38.
\item
Pasche, O.C., Lam, H., and Engelke, S. (2025) ``Extreme Conformal Prediction: Reliable Intervals for High-Impact Events.'' arXiv.2505.08578v1 [stat.ME]
\item
Pathak, J., Lu, Z., Hunt, B. R., Girvan, M., and Ott, E. (2022). ``Using Machine Learning to Improve Weather Forecasting.'' \textit{Science} 377(6609):  1111 --1115.
\item
Peng, S., Piao, S., Ciais, P. et al.  ( 2013)``Asymmetric Effects of Daytime and Night-time Warming on Northern Hemisphere Vegetation. Nature 501(1243): 88 -- 92.
 \item
Perkins, S.E. (2015) ``A review on the Scientific Understanding of Heatwaves --Their
Measurement, Driving Mechanisms, and Changes at the Global Scale.'' \textit{Atmospheric Research} 165 - 165: 242 -- 267.
\item
Perkins, S.E., and Alexander, L.V., (2013) ``On the Measurement of Heat Waves.'' \textit{Journal of Climate} 26: 4500 -- 4517.
\item    
Petoukhov, V., Rahmstorf, S., Petri, S., and Schellnhuber, H.J. (2013) ``Quasiresonant Amplification of Planetary waves and Recent Northern Hemisphere Weather Extremes.'' \textit{Proceedings of the National Academy of Sciences} 110: 5336 -- 5341.
\item
Pitcar, A., Cheval, S., and Frighenciu, M. (2019) ``A Review of Recent Studies on Heat wave Definitions, Mechanisms, Changes, and Impact on Mortality.'' \textit{Forum Geographic} XVII (2): 103 -- 120.
\item
Price, I., Sanchez-Gonzalez, A., Alet, F. et al. (2024) ``Probabilistic Weather Forecasting with Machine Learning.'' \textit{Nature} 624: 559 -- 563.
\item
Ridgeway. G., (2024) ``Generalized Boosted Models: A Guide to The \textit{gbm} Package.'' \newline https://cran.r-project.org/web/packages/gbm/vignettes/gbm.pdf
\item
Romano, Y., Patterson, E., and Cand\`{e}s, E.J. (2019) ``Conformal Quantile Regression.'' In H. Wallach et al., (eds) \textit{Advances in Neural Information Processing Systems}, Volume 32.
\item
Russo, S., Dosio, A., Graverson, R.G., Sillmann, J., Carrao, H., Dunbar, M.B., Singleton, A., Montanga, P., Barbola, P., and Vogt, J. (2014) ``Magnitude of Extreme Heat Waves in Present Climate and Their Projection in a Warming World.'' \textit{Journal of Geophysical Research Atmospheres} 199(22) 12,500 -- 12,512.
\item
Smith, T.T., Zaitchik,  B.F., Gohlke, J.M. (2013) ``Heat Waves in the United States: Definitions, Patterns and Trends.'' \textit{Climatic Change} 118: 811 -- 825.
\item
Steadman, R.G., (1984( ``A Universal Scale of Apparent Temperature.'' \textit{Journal of Applied Meteorology and Climatology} 23(12): 1674 -- 1687.
\item
Stull, R., (2017) \textit{Practical Meteorology: An Algebra-based Survey of Atmospheric Science} - version 1.02b.  University of British Columbia. 
\item
Velthoen, J., Dombry, C., Cai, JJ., and Engelke,  S. (2023) ``Gradient Boosting for Extreme Quantile Regression.'' \textit{Extremes} 26: 639 -- 667.
\item
Warner, T.T. (2004) \textit{Desert Meteorology} Cambridge University Press.
\end{description}
\end{document}